\documentclass[twocolumn,aps,prb,epsf,graphics,psfig]{revtex4}
\usepackage{graphicx}

\begin{document}
\title{
Computational studies for reduced graphene oxide in hydrogen-rich environment
}

\author{Ramin M. Abolfath$^{1,2}$, Kyeongjae Cho$^2$}

\affiliation{
$^1$School of Natural Sciences and Mathematics, University of Texas at Dallas, Richardson, TX 75080\\
$^2$Department of Materials Science, University of Texas at Dallas, Richardson, TX 75080
}

\date{\today}

\begin{abstract}
We employ molecular dynamic simulations to study the reduction process of graphene-oxide (GO) in a chemically active environment enriched with hydrogen. We examine the concentration and pressure of hydrogen gas as a function of temperature in which abstraction of oxygen is possible with minimum damage to C-sp$^2$ bonds hence preserving the integrity of the graphene sheet. Through these studies we find chemical pathways that demonstrate beneficiary mechanisms for the quality of graphene including formation of water as well as suppression of carbonyl pair holes in favor of hydroxyl and epoxy formation facilitated by hydrogen gas in the environment.
\end{abstract}
\maketitle

\section{Introduction}
Following the progress in the fabrication and mechanical exfoliation of the
graphene~\cite{Novoselov2004:S,Novoselov2005:N,Zhang2005:N,Zhou2006:N},
extraordinary low-dimensional electrical and mechanical properties of graphene have been
revealed~\cite{Castro-Neto2009:RMP}.
Currently, mass-scale fabrication of graphene-based devices free from defects and imperfections requires chemical processing with high quality product and is one of the challenging technological problems for graphene device technology. For example, the micro-mechanical cleavage of graphite~\cite{Novoselov2005:PNAS}, epitaxial growth on silicon carbide~\cite{Berger2004:JPCB}, chemical vapor deposition of hydrocarbons on transition metal surfaces~\cite{Kim2009:Nature} have shown difficulties in obtaining processable graphene sheets in large quantities thus impeding full exploitation of its exciting properties.
Chemical oxidation of graphite and subsequent exfoliation in solution allow a large scale production of isolated graphene oxide (GO). However, the reduction of GO is known to create many structural defects degrading the material properties of reduced GO.
Hence identifying reliable methods that allow removal of oxygen from GO with minimum damage to C-sp$^2$ bonds in graphene-structures is the focus of current research activities.

Recently studies have shown that the chemical reduction of graphite oxide can be a low-cost and scalable method~\cite{Park2009:NT,Stankovich2006:N,Stankovich2006:JMC}. It has been found that layered materials constituting graphene layers functionalized with epoxy and hydroxyl groups are easily exfoliated in water. The resulting graphene oxide (GO) monolayers can be deposited in controllable density onto a large variety of substrates, thus enabling the preparation of thin conductive films on solid and flexible substrates~\cite{Wang2008:NL,Eda2008:NN,Li2008:NN}.

It is accepted that GO can be described as a random distribution of oxidized areas with the oxygenated functional groups, combined with nonoxidized regions wherein most of the carbon atoms preserve sp$^2$ hybridization~\cite{Lerf1998:JPCB}.
GO is electrically insulating material, however, its conductivity can improve up to 4 orders of magnitude by chemical reduction~\cite{Jung2008:NL,Gilje2007:NL,Gomez-Navarro2007:NL}, nevertheless the typical conductivities of reduced GO (RGO) is still lower than pristine graphene by a factor of 10-100~\cite{Lopez2009:AM,Tung2009:NN} due to presence of residual functional groups remaining after reduction. In contrast to mechanically exfoliated graphene, the chemically derived graphene (i.e., RGO) is found to contain a considerable amount of topological defects.
Similar to other low-dimensional carbon nanostructures like carbon nanotubes~\cite{Gomez-Navarro2005:NM}, graphene ribbons~\cite{Nakada1996:PRB,Wakabayashi1999:PRB,Wunsch2008:PRL}, fullerenes~\cite{Saito1992:CPL} and more recently graphene quantum dots~\cite{Bittner2010:PRB}, topological defects and etch-holes (that are unavoidable products of the reduction process) are expected to strongly affect the device electronic and mechanical performance, and thus to account for the differences between RGO and pristine graphene.

Recent advances in computational modeling of materials such as molecular dynamic (MD) simulations and ab-initio calculation have enabled researchers to provide detailed studies in the microscopic level and allows a careful analysis of the chemical reactions and diffusion mechanisms~\cite{Acik2010:AN,Bagri2010:JPC,Bagri2010:NC,Cai2008:S,Zabo2006:CM,He1998:CPL,Paci2007:JPC,Park2009:NN,Bourlinos2003:L,Gao2009:NM,Wang2010:PRB,Wang2010:AN,Wang2011:PCCP,Fonseca2011:PRB,vanDuin2001:JPC,Abolfath2011:JPC}.
Such studies are helpful in identifying the experimental conditions that allows the minimization of mechanical and chemical damages to graphene and the optimization of their performance~\cite{Acik2010:AN,Bagri2010:JPC}.

In this work we employ MD simulations based on ab-initio CPMD~\cite{CPMD} and ReaxFF~\cite{vanDuin2001:JPC,Abolfath2011:JPC} which are the mathematical formulation that governs the appropriate dynamics of the molecular system to analyze the carbon-oxygen chemical reaction and reduction process.
We show that the GO structural relaxation assisted with the oxygen surface diffusion that allows the migration of epoxy functional groups and protects the C-sp$^2$ bonds, preserve the integrity of the graphene lattice structure and governs the dynamical pathway in low temperatures. In contrast, in high temperatures, consistent with recent finding of Ref.~\cite{Acik2010:AN,Bagri2010:JPC}, we show that the formation of carbonyl is dynamically favorable and can be accounted for dominant mechanism in breaking C-sp$^2$ bonds.
We then consider immersed systems of GO in dilute gas of hydrogen using two large-scale structural models for GO and demonstrate the effect of hydrogen in removal of oxygen, blocking the carbonyl pair formation and finally seek for conditions in which structural damage to C-sp$^2$ bonds is minimized.

\section{MODEL AND COMPUTATIONAL METHODS}
The atomic model structure of large scale GO is constructed by including randomly distributed hydroxyl and epoxy functional groups to a single-graphene sheet.
The dynamical stability of the entire structure in particular the functional groups attached to the graphene will be investigated by performing MD in which the time evolution of the system and its environment is calculated.
The system consists of GO in vacuum or immersed in hydrogen gas can exchange energy and particles with the environment.
In vacuum the exchange of energy is possible by contacting GO with a canonical thermostat that keeps temperature constant after reaching the thermodynamic equilibrium. In this case the exchange of particles is possible either if the functional groups in GO are in a highly unstable configuration or if they undergo a chemical reaction and form chemical complexes. If the molecules formed by such complexes are stable they can be detached from their host and diffuse thermally away from GO.
In an environment rich with hydrogen molecules we further expect that the reaction-diffusion processes are accounted for the particle exchange between GO and its environment.

Our ab-initio/DFT calculation consists of Car-Parrinello molecular dynamics (CPMD) model~\cite{CPMD}, in which the potential energy of the system can be calculated on the fly, as needed for the conformations of the dynamical trajectory to simulate the chemical reaction pathways.
It is implemented in a plane-wave basis within local spin density approximation (LSDA) with an energy
cutoff of 70 Ry (rydberg), and with Becke~\cite{Becke1988:PRA} exchange and Lee-Yang-Parr (BLYP) gradient-corrected functional~\cite{Lee1988:PRB}. Norm conserving ultrasoft Vanderbilt pseudopotentials were used for
oxygen, hydrogen, and carbon. The CPMD microcanonical dynamics (constant energy ensemble) that allows achieving dynamical equilibration for $T=$300-1500 K were performed, after wave function optimization. Then a requenching of the wave function and canonical CPMD is performed. A supercell with periodic boundary condition in the plane of graphene  with Poisson solver of Martyna and Tuckerman was used~\cite{Martyna1999:JCP}.

Our approach for reactive MD consists of ReaxFF~\cite{vanDuin2001:JPC} implemented in Large-scale Atomic/Molecular Massively Parallel Simulator (LAMMPS)~\cite{LAMMPS}. The consistency and accuracy of ReaxFF with DFT calculation is checked systematically by employing ab-initio CPMD~\cite{CPMD}.
In ReaxFF the atomic interactions are described by the reactive force field potential~\cite{vanDuin2001:JPC}. ReaxFF is a general bond-order dependent potential that provides accurate descriptions of bond breaking and bond formation. Recent simulations on a number of hydrocarbon-oxygen systems~\cite{vanDuin2001:JPC}, GO~\cite{Acik2010:AN,Bagri2010:JPC,Bagri2010:NC}, and organic molecules~\cite{Abolfath2011:JPC} showed that ReaxFF reliably provides energies, transition states, reaction pathways and reactivity trends in agreement with ab-initio calculations and experiments.
The information on the type of chemical reactions and the time-evolution of the damage is collected via running the MD up to 5 ps where the rearrangement of the atomic coordinates have been deduced from a dynamical trajectory calculated by ReaxFF. These simulations performed using periodic boundary conditions in a canonical NPT ensemble with a Nose-Hoover thermostat for temperature control and a time step of 0.25 fs.
Finally a fitting to non-linear reaction-diffusion process is performed.

\begin{figure}
\begin{center}
\includegraphics[width=0.7\linewidth]{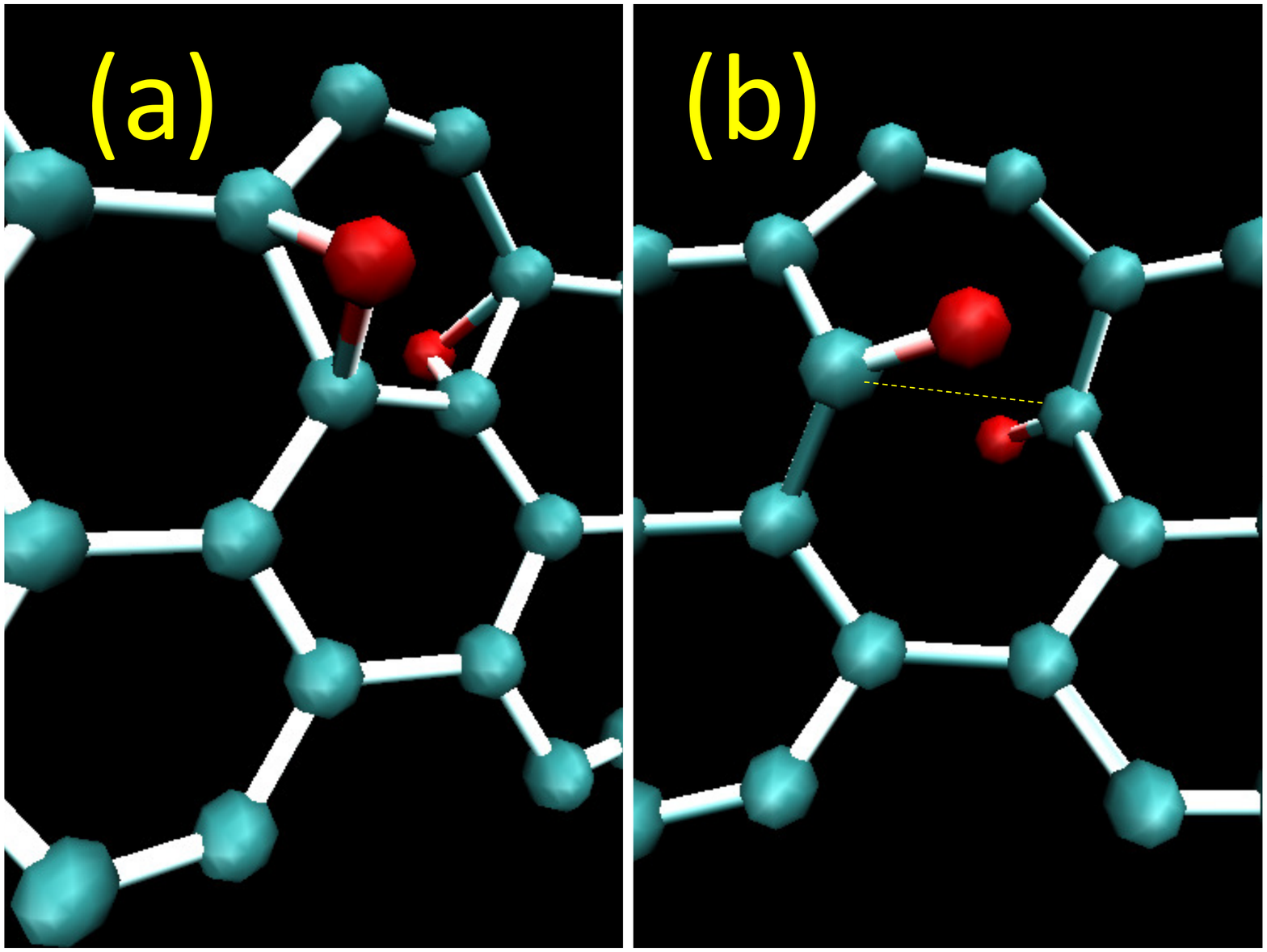}\vspace{-0.5cm}
\includegraphics[width=0.9\linewidth]{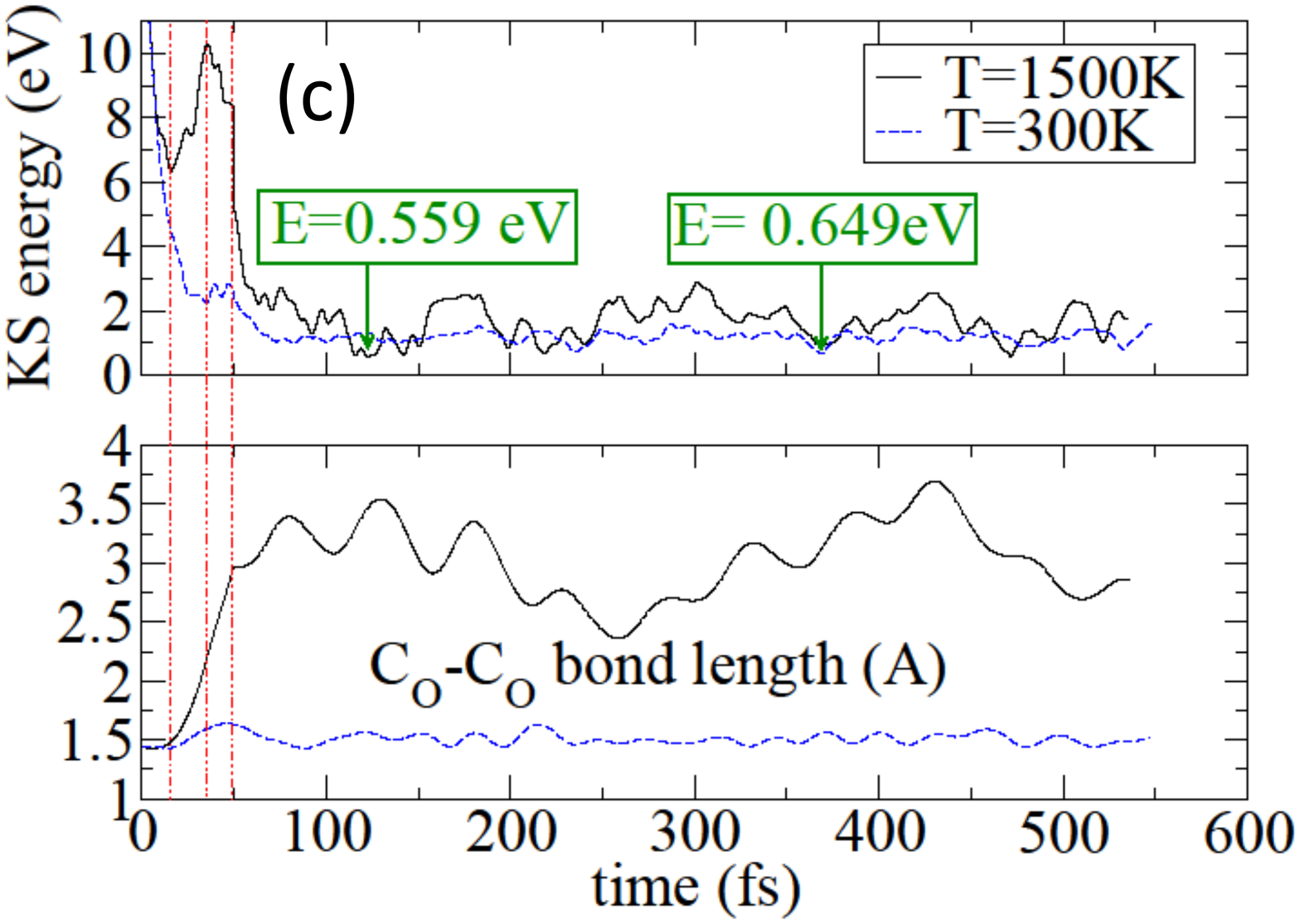}
\noindent
\caption{
Epoxy and carbonyl pair formation of GO at $T=300$K (a) and $T=1500$K (b).
(c) Kohn-Sham energy and C-C bond length (in carbonyl groups, shown by a dash-line in figure b) as a function of time and temperature. The transition state for C-C bond breaking corresponding to the energy barrier $\Delta \approx 3$eV shown between the first two dash lines. A drop in Kohn-Sham energy is an indication of formation of carbonyl groups. The thermal fluctuations in C-C bond length $d\approx 3\AA$ predicted by ab-initio CPMD shows the stability of carbonyl pair.
The lowest energy difference between the configurations in $1500$K and $300$K corresponding to pair of carbonyls and epoxys is calculated 0.09 eV within the present MD time.
This small energy difference (almost one order of magnitude smaller than the one reported in Ref.~\cite{Bagri2010:NC}) is due to relaxation of the structural strain by the surface diffusion of epoxys.
Carbon and oxygen atoms are shown as green and red.
}
\label{fig1}
\end{center}\vspace{-0.5cm}
\end{figure}

\section{RESULTS AND DISCUSSIONS}
\subsection{Carbonyl vs. epoxy formation, DFT results}
First we present series of ab-initio CPMD calculation to study the oxygen-graphene interaction as a function of temperature.
At room temperature $T=300$K, Fig.~\ref{fig1}(a) shows formation of epoxy groups by two oxygen initially located in the opposite sides of the graphene sheet where carbon and oxygen atoms are shown with green and red balls.
Increasing temperature up to $T=1500$K, as shown in Fig.~\ref{fig1}(b), change the pathway to carbonyl formation that accompany with formation of holes.
Figs.~\ref{fig1}(c) and (d) show the respective Kohn-Sham (KS) energies equivalent to potential energies in classical MD  and carbon-carbon bond length (shown by arrow in Figs.~\ref{fig1}(c) and (d)). The simulation is performed up to $t=600$ fs where it shows stability of the system under study.
Accordingly the transition for breaking C-C bond requires passing through an energy barrier $\Delta \approx 3$eV. In high temperatures the shear force induced by anti-parallel and out-of-plane vibrational modes of oxygen-carbon bonds acting on carbon-carbon bonds provide enough kinetic energy that allows passing through the transition state. In this case the in-plane C-sp$^2$ bond breaks and the formation of carbonyl pair lowers the local strain energy of graphene.

In lower temperatures, however, the vibrational energy is not large enough to break the C-C bonds. Instead, system finds another pathway to a stable configuration by moving two oxygen one lattice constant away from each other such that it lowers the planar strain energy. This relaxation of local energy is assisted by the surface diffusion of the oxygen pairs.
Because the chemical bond breaking is much faster process than mechanical expansion of the graphene lattice structure, the structural relaxation in achieving the minimum energy configuration for a carbonyl pair occurs faster than epoxy pair. Within 0.6 ps of CPMD time, we find the energy difference between two lowest-energy configurations of carbonyl and epoxy pairs $\Delta E_{\rm KS}\approx -0.1$ eV. However, a small and steady negative-slope seen in KS energy of epoxy pairs implies that the energy of epoxy pair asymptotically falls to a lower energy than the energy of carbonyl-pair.

As the oxygenation of the graphene involves with passing through a large energy barrier in the reaction-diffusion processes, and because the ground state energy of both carbonyl and epoxy configurations are approximately degenerate,
the carbonyl formation is expected to be dominant at high temperature where the vibrational modes of the oxygen-carbon bonds fluctuate strongly and thus oxygen becomes more damaging whereas in lower temperature the dominant surface diffusion allows stability of the epoxy groups without forming the structural holes.
In the latter case the local strain energy can be relaxed through the thermal expansion of the lattice assisted by in-plane phonon modes.

\begin{figure}
\begin{center}
\includegraphics[width=0.7\linewidth]{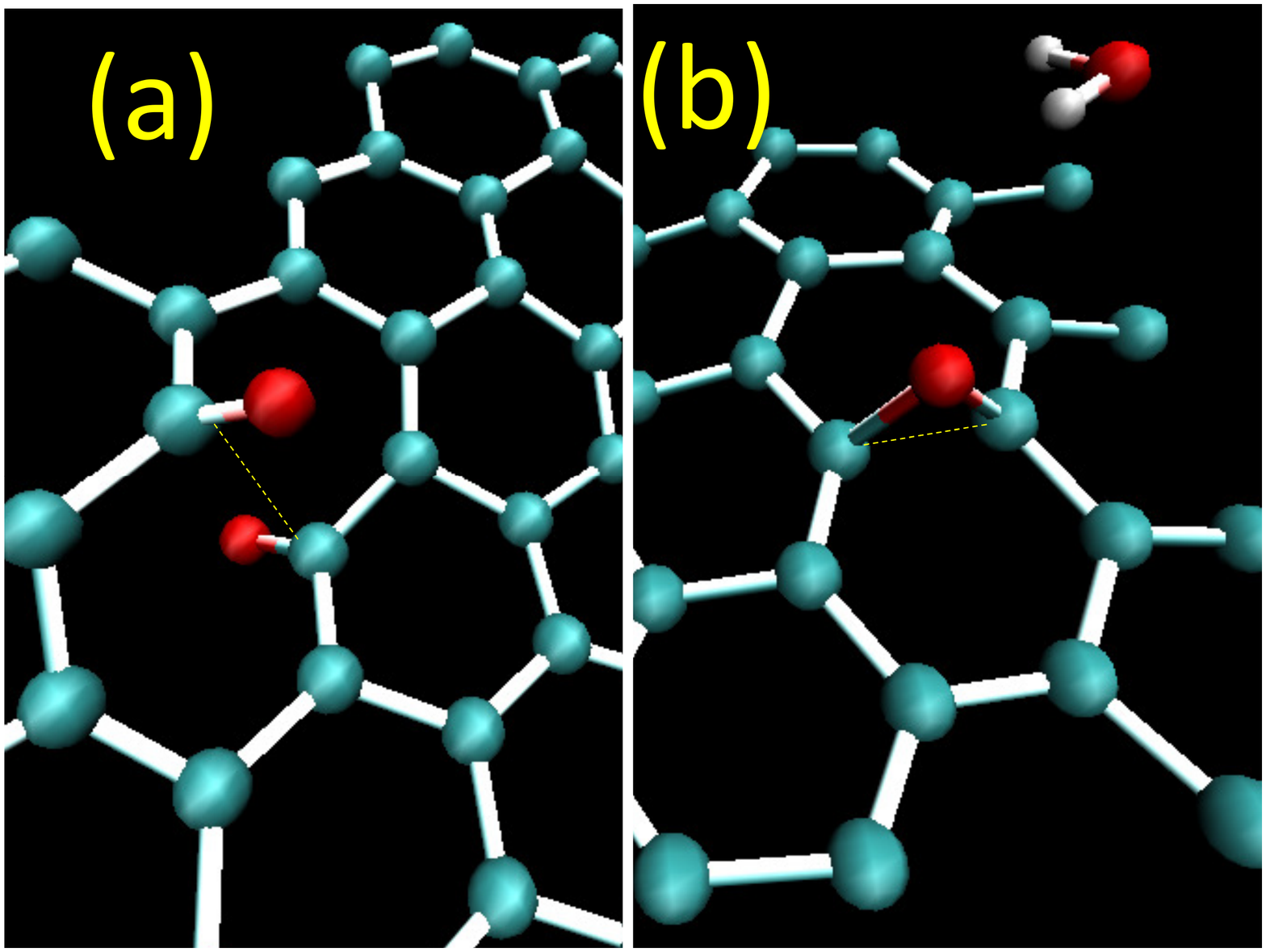} \vspace{-0.5cm}
\includegraphics[width=0.9\linewidth]{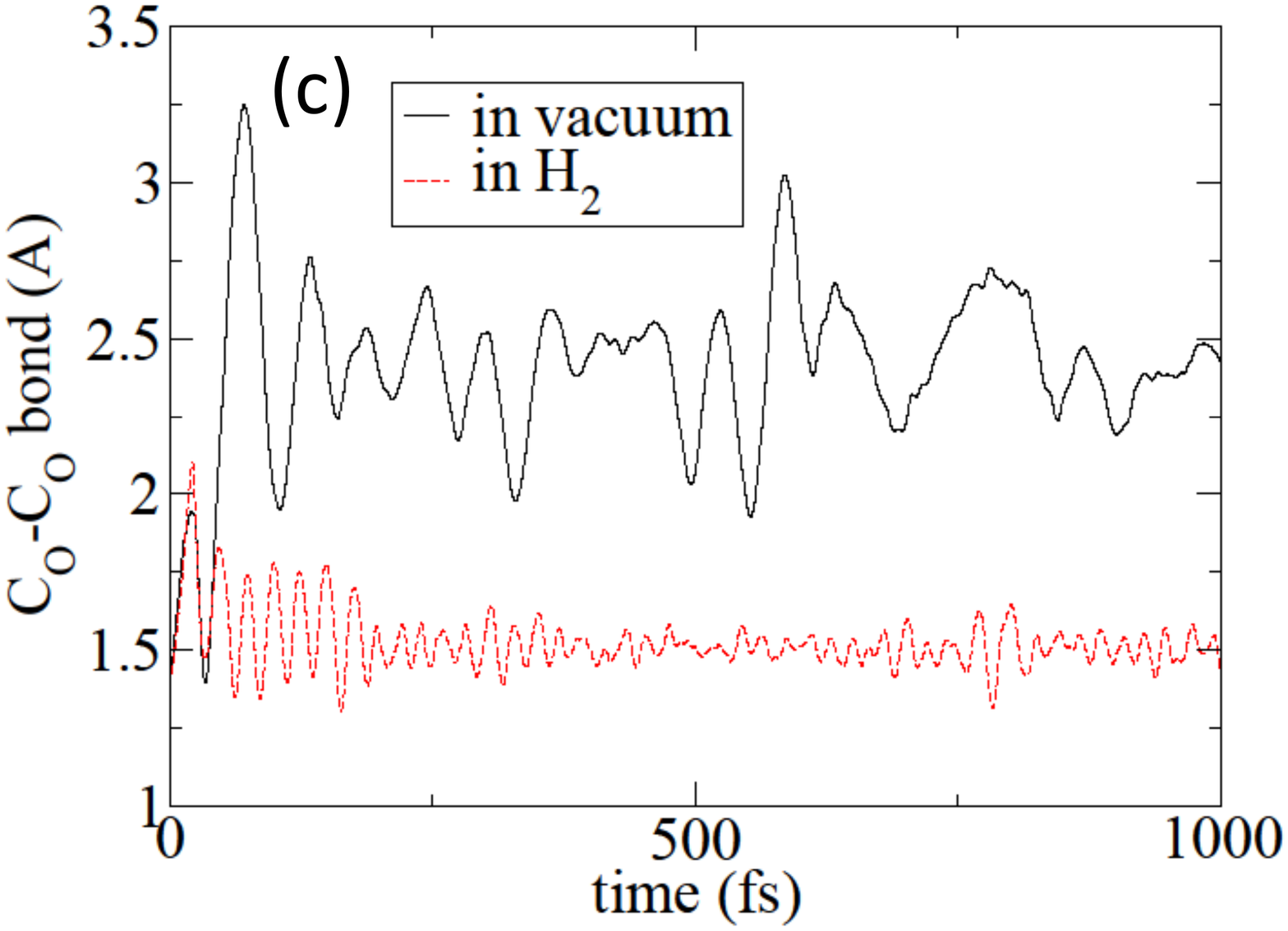}
\noindent
\caption{
(a) Carbonyl formation at $T=1500$K using ReaxFF-MD.
(b) Similar simulation in $H_2$ rich envoronment. The final products are water molecule and epoxy group.
(c) C-C bond length (shown by dash line in figures a and b) as a function of time calculated at $T=1500$K in vacuum and in $H_2$ environment. Carbon, oxygen and hydrogen atoms are shown as green, red and white, respectively.
}
\label{fig2}
\end{center}\vspace{-0.5cm}
\end{figure}

\begin{figure}
\begin{center}
\includegraphics[width=0.9\linewidth]{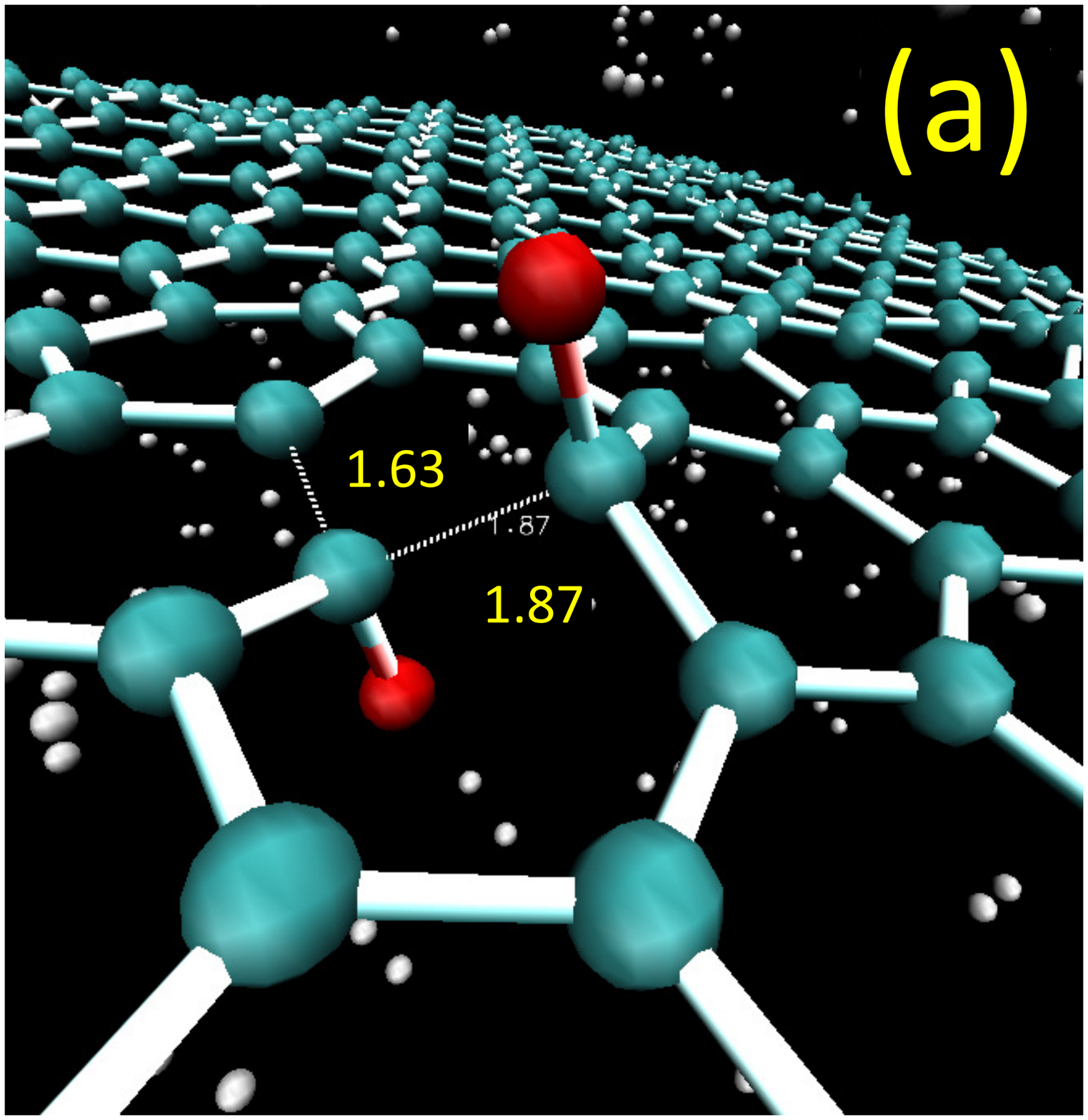}
\includegraphics[width=0.9\linewidth]{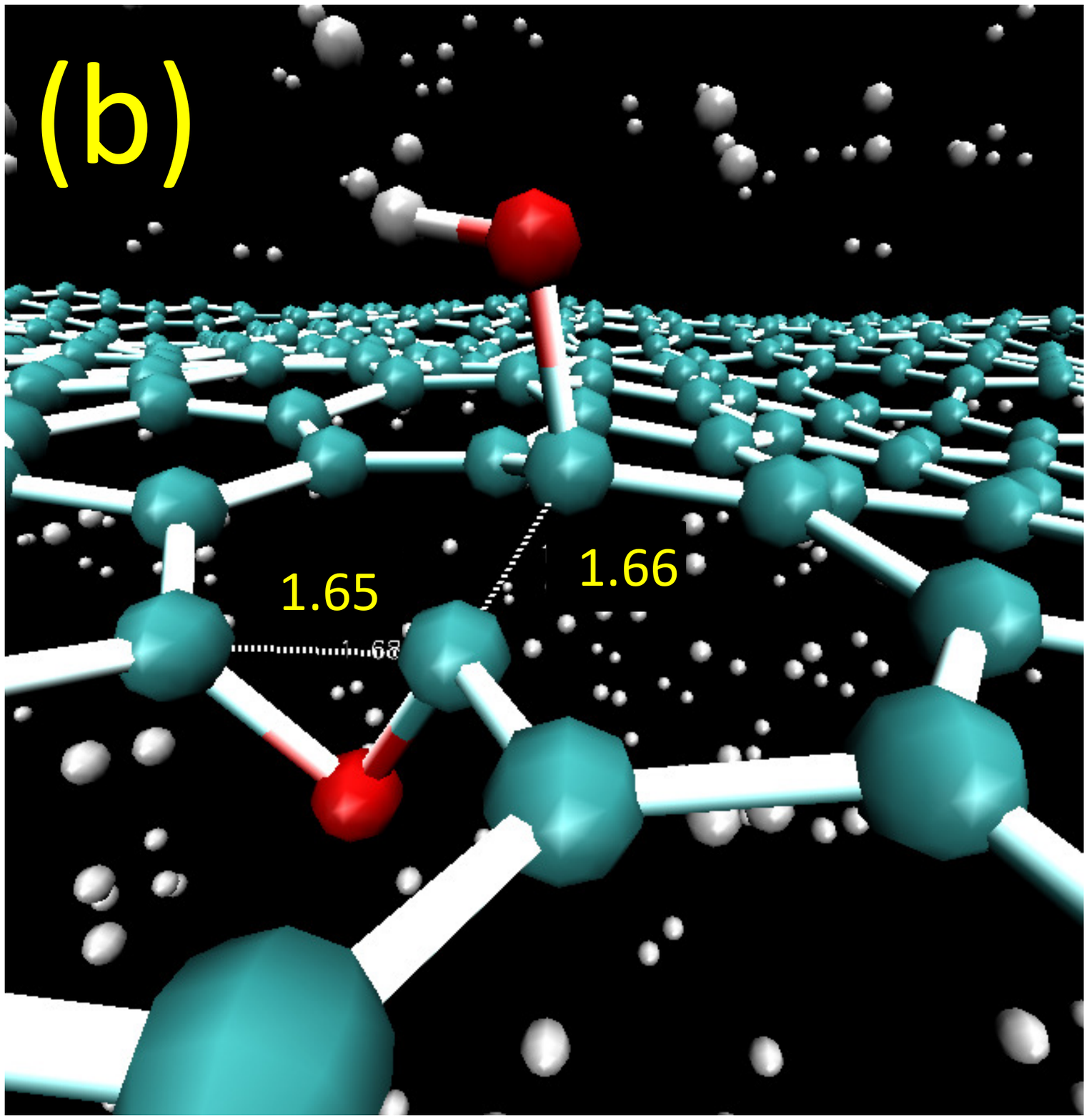}
\includegraphics[width=0.9\linewidth]{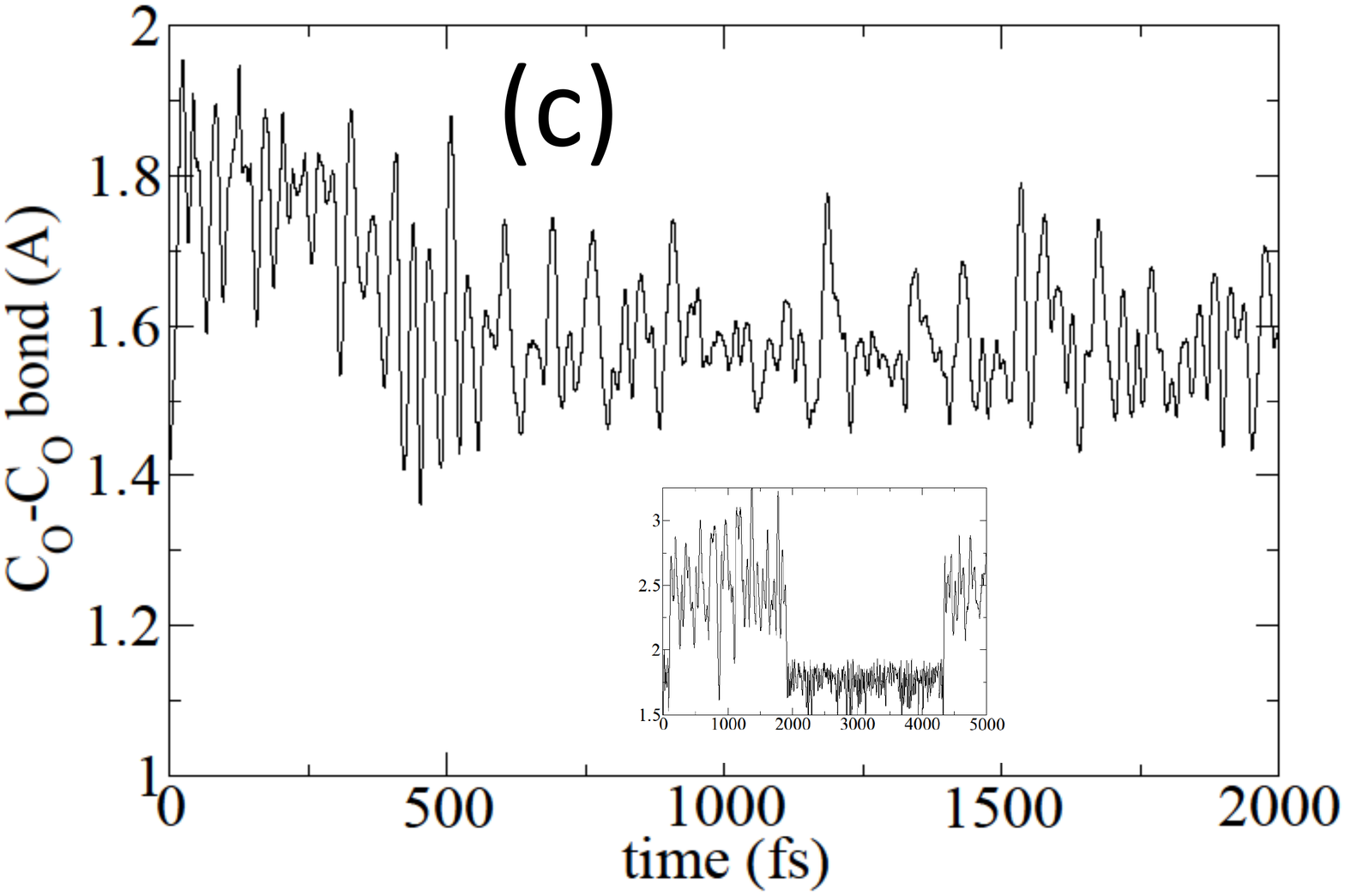}
\noindent
\caption{
(a) An intermediate configuration corresponding to the onset of carbonyl formation found in the hydrogen-rich environment. This configuration shows stability up to $t\approx 500$ fs using ReaxFF-MD. Here two oxygen are initially located in the opposite sides of the graphene sheet and the temperature is $T=1500$ K.
(b) Final state of the same ReaxFF-MD that shows formation of hydroxyl and epoxy functional groups for $t > 500$ fs. This chemical pathway shows suppression of carbonyl pair in favor of formation of hydroxyl and epoxy functional groups in the hydrogen-rich environment at $t\approx 500$ fs. Two dash-lines used to show the broken carbon-carbon bond length.
(Bottom) C-C bond length as a function of time. The spike at the beginning is an indication of the onset of carbonyl bond formation that is suppressed due to the presence of hydrogen molecules (seen as white spheres) and the transformation to hydroxyl and epoxy functional groups with $d_{CC}\approx 1.6 \AA$.
Shown in inset is the same calculation in vacuum performed up to $t=5$ ps. It shows a bi-stability behavior in the equilibrium position of carbonyl-pair distance that jumps between 1.8$\AA$ and 2.5$\AA$.
}
\label{fig2_2}
\end{center}\vspace{-0.5cm}
\end{figure}

\subsection{Carbonyl formation, ReaxFF results}
The above analysis, consistent with the empirical data~\cite{Bagri2010:NC}, indicates that formation of the carbonyl holes in high temperature annealing processes can not be avoided.
A post-processing in a chemically active environment that allows removal of oxygen is an alternative possibility that may lower the chance in carbonyl formation as well as reducing the oxygen concentration in GO.
In this section, we examine such possibility.

In Fig.~\ref{fig2}(a) we describe a pathway in carbonyl pair formation in vacuum obtained after 1ps MD simulation by ReaxFF. To check the consistency of the results with ab-initio CPMD, presented in previous section, we start with
an initial configuration that consists of two oxygen located at the opposite side of graphene with separation of one lattice constant.
Fig.~\ref{fig2}(b) shows result of a similar MD simulation starting from the same initial configuration used for Fig.~\ref{fig2}(a), except that hydrogen molecules are added in the computational box. Here carbon, oxygen and hydrogen atoms are shown as green, red and white balls respectively.

We first energy minimize the system of hydrogen and leave a separation gap between GO and hydrogen gas to avoid any bad contact and check the convergence of the results as a function of gap size to minimize the computational artifact introduced by the separation gap. As it is shown, a water molecule forms after removal of one oxygen by hydrogen (for pictorial clarity, other hydrogen molecules are not shown). The second oxygen converts to an epoxy, hence the hole associated to carbonyl pair does not form.
The bold-line in Fig.~\ref{fig2}(c) shows the corresponding carbon-carbon distance, $d_{CC}$, as a function of time in vacuum. At $t=0$ it starts from graphene lattice constant $d_{CC}=1.42\AA$ and approximately at $t=50$ fs it undergoes strong fluctuations and jumps to $d_{CC}\approx 2.5\AA$.
In Fig.~\ref{fig2}(c), the dash-line shows the time-evolution of C-C distance in the hydrogen-rich environment. We observe the onset of carbonyl formation at the beginning indicated by a peak in the amplitude of C-C distance oscillations, however it is immediately suppressed by oxygen-hydrogen chemical interaction that leads to water formation.

Similar calculation based on ReaxFF-MD starting from slightly different initial condition is shown in Fig.~\ref{fig2_2}. It illustrates an intermediate configuration and formation of carbonyl pair (Fig.~\ref{fig2_2}a) within $t\leq 300$ fs in which an absorption of hydrogen by oxygen occurs. This process converts one of the carbonyl pair oxygens to hydroxyl. Around $t\approx 600$ fs the second oxygen forms a bond with a neighboring carbon and an epoxy functional group is formed. This is shown in Fig.~\ref{fig2_2}b.
In Fig~\ref{fig2_2}(c), the time evolution of C$_{\rm O}$-C$_{\rm O}$ bond length, corresponding to the dash line in the right side of Fig.~\ref{fig2_2}(a), shows gradual contraction, a manifestation on carbonyl pair suppression.

To check the establishment of carbonyl pair in vacuum for this particular simulation, we performed ReaxFF-MD up to 5 ps where we observed a bi-stability behavior in the equilibrium position of C$_{\rm O}$-C$_{\rm O}$ distance that jumps between 1.8$\AA$ and 2.5$\AA$, seen in the inset of Fig~\ref{fig2_2}(c). Note that $d_{CC}\approx 3\AA$ is predicted by ab-initio CPMD (see Fig.~\ref{fig1}(c)). In vacuum, we observe that the onset of transition to C$_{\rm O}$-C$_{\rm O}$ distance with 2.5$\AA$ (carbonyl formation) occurs around $t\approx 100$ fs. This is in contrast with the same calculation in the hydrogen-rich environment, in which C$_{\rm O}$-C$_{\rm O}$ distance gradually decays to 1.6$\AA$ (without forming carbonyl pair) until a hydroxyl functional group forms around $t\approx 300$fs. This comparison shows that within $100 \leq t \leq 300$ fs, hydrogen-molecules in atmosphere hinder the carbonyl formation, by applying a physical pressure to epoxy pair without forming a chemical reaction.

\begin{figure}
\begin{center}
\includegraphics[width=0.6\linewidth]{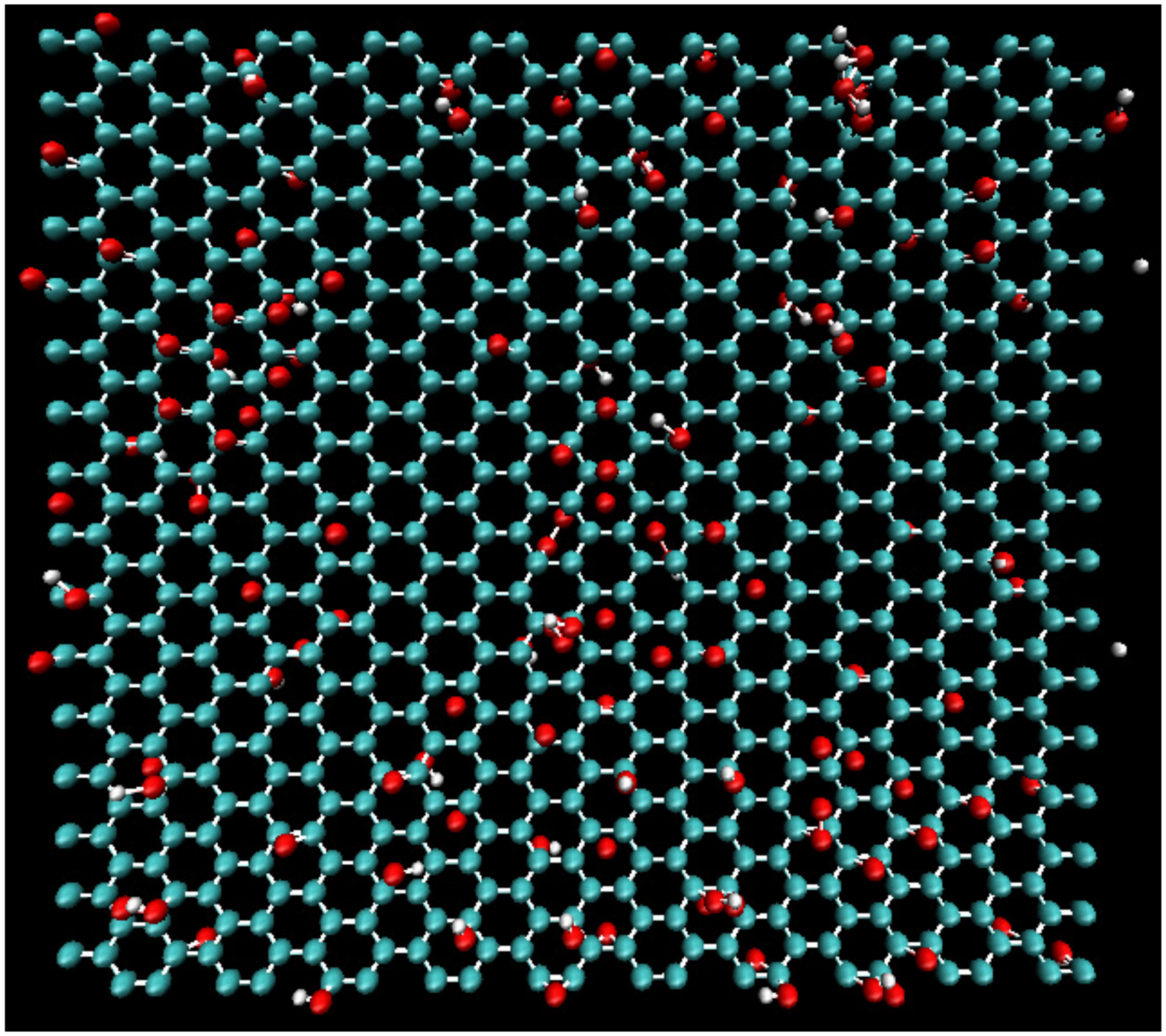}\\ \vspace{0.5cm}
\includegraphics[width=0.6\linewidth]{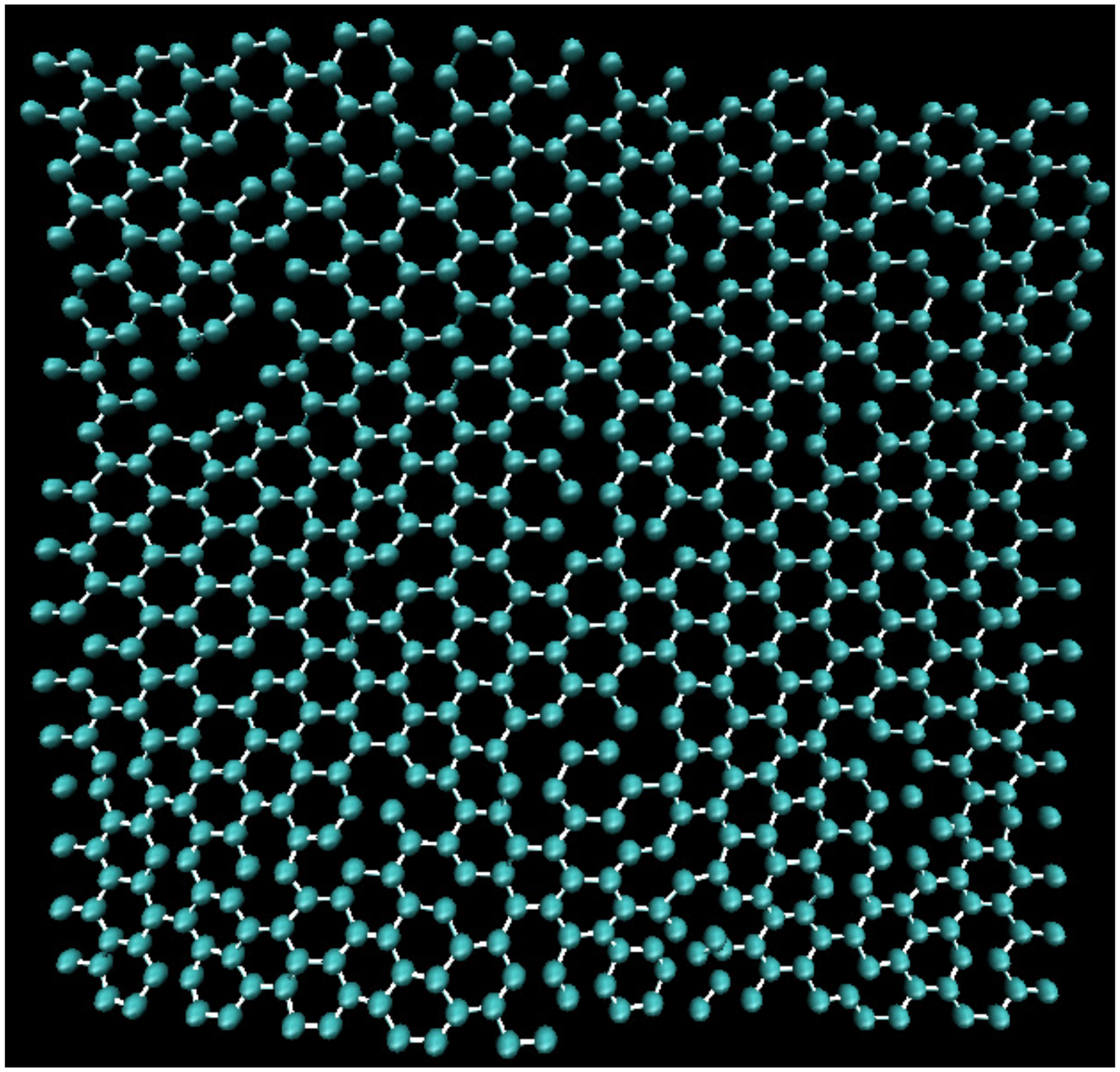}\\ \vspace{0.5cm}
\includegraphics[width=0.6\linewidth]{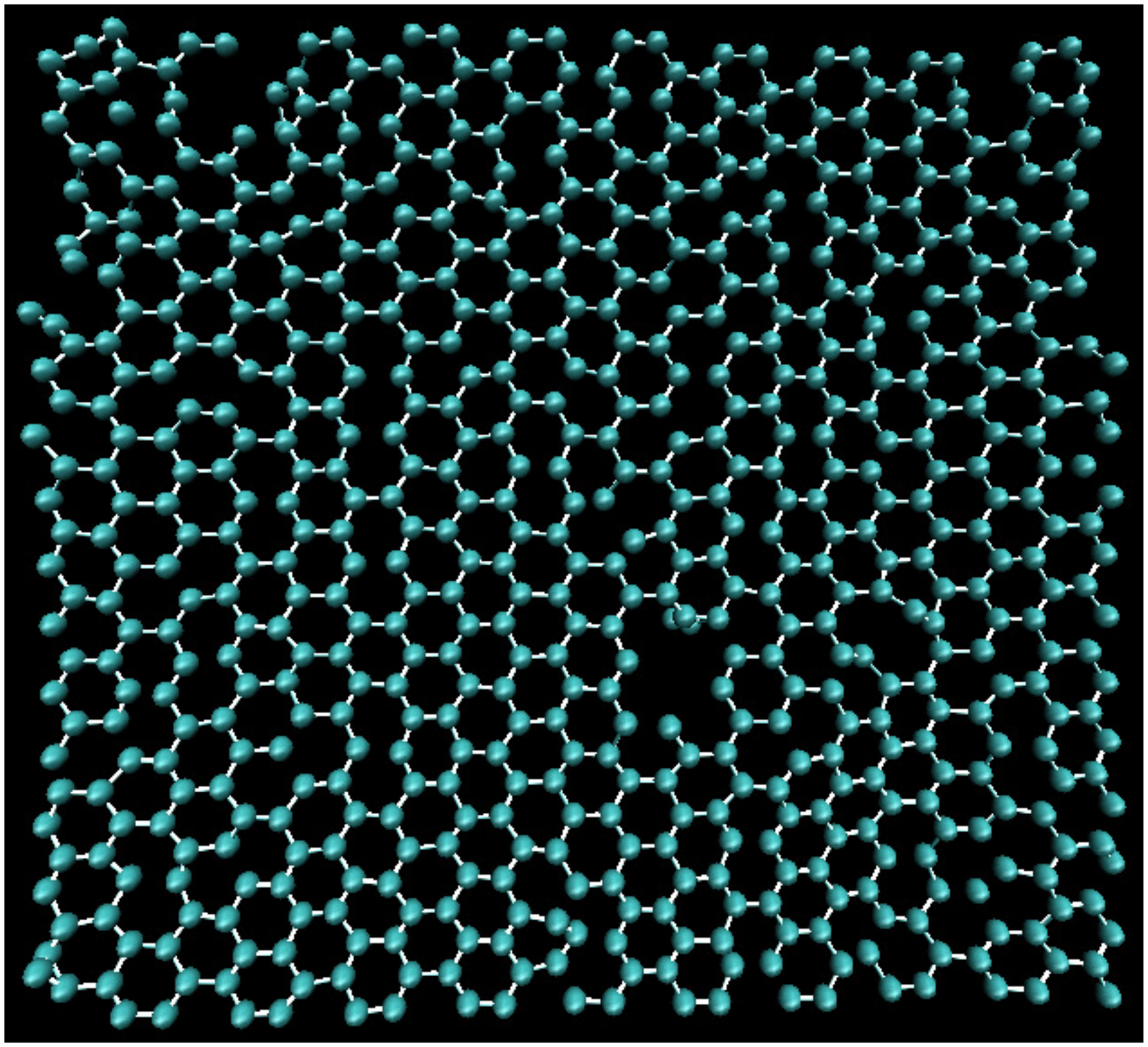}\\ \vspace{0.5cm}
\noindent
\caption{
Initial (left) and final states of GO after annealing at $t=2$ps, and $T=1500$K in vacuum (middle) and in $H_2$-rich environment with the density of $n_{\rm H}\approx 0.24 \times 10^{23} {\rm cm}^{-3}$ (right). For clarity in visualization only carbons are shown in the final states. The holes in GO are due to formation carbonyl and other functional groups.
}
\label{fig2a}
\end{center}\vspace{-0.5cm}
\end{figure}

We now turn to a system with large number of epoxy and hydroxyl functional groups intended to be a model for realistic structure of GO.
Following Bagri {\em et al.}~\cite{Bagri2010:JPC,Bagri2010:NC}, we start from an initial GO where the position of carbons follow the long-range order structure of pristine graphene. Fig.~\ref{fig2a} shows the initial and final structures in vacuum and in hydrogen-rich environment. The final structures are shown after 1ps MD run.
For clear visualization of the carbon structures, we omitted oxygens and hydrogens in the final configurations.
We initially distribute oxygens and hydrogens randomly as epoxy and hydroxyl groups on two sides of graphene sheet with approximate area coverage of 12\% epoxy and 7\% hydroxyl using periodic boundary condition. Here the GO consists of $N_{\rm C}=640$, $N_{\rm O}=122$, and $N_{\rm OH}=43$, where 79 oxygens are in form of epoxy and the rest are hydroxyl groups.

\begin{figure}
\begin{center}
\includegraphics[width=1.0\linewidth]{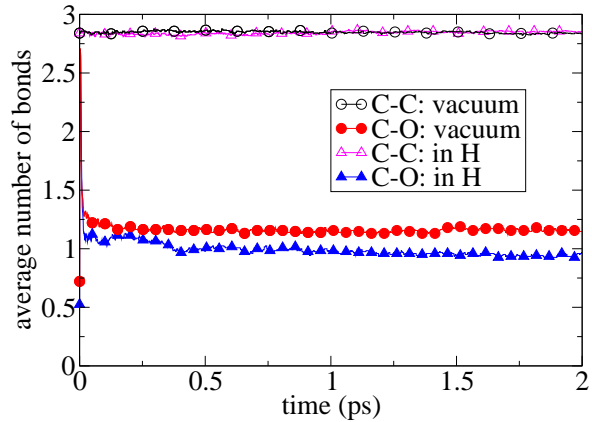}\\ \vspace{0.5cm}
\noindent
\caption{
Time evolution of the average coordinate number of carbon and oxygen calculated as average bond between two carbons (C-C) and carbon-oxygen (C-O). A cut-off distance $d^*=2\AA$ is used to count the number of separated atoms from GO.
}
\label{fig2c}
\end{center}\vspace{-0.5cm}
\end{figure}

The number of C-sp$^2$ bonds per area is in particular the main factor that determines the quality of the graphene structure.
These defects observed in GO, are absent in pristine graphene where such planar distortion does not occur and carbon coordination number does not change during MD by Hydrogen gas in environment.

The number of carbons and oxygens bounded in graphene-sheet can be calculated as a function of time by counting the number of C-C and C-O bonds per carbon and oxygen atom respectively.
First we note that we found no chemically-unbounded carbon from the planar GO either by thermal annealing or by hydrogen gas. Therefore we discard the possibility in the chemical removal of carbon from GO.
The average number of carbon-bonds can be considered as representation of the carbon coordination number in the graphene-sheet. During the annealing process, the epoxy and hydroxyl groups initially localized in the GO may interact and form free water molecules, OH free radicals, oxygen, ozone and hydronium molecules ($H_3O$). To count the number of abstraced oxygens, we calculate the distance between oxygens and carbons, $d_{oc}$. We define a length-scale as a cut-off such that if $d_{oc} \geq d^*=2\AA$ the chemical bonds is accounted for the broken-bonds. The result of this calculation shown in Fig.~\ref{fig2c} indicates that in vacuum the
average C-O bonds reach steady state shortly after starting MD, however, in hydrogen-rich environment it decreases gradually as a function of time.
We therefore conclude that hydrogen in environment do not suppress C-C bond breaking (e.g., etch hole formation), but that they facilitate continuous removal of oxygen leading to reduced C-O bonding in GO.
This is consistent with the expectation on the role of hydrogen in GO reduction.

Some interesting events responsible for the structural improvement of graphene sheet and observed in our simulation include the collision between hydrogen molecules and oxygen functional groups that lead to capturing of hydrogen by oxygen and formation of a hydroxyl group on the surface of GO and finally formation of free water and hydronium molecules.
In high temperatures where the average kinetic energy of hydrogen molecules is larger, this happens more frequently and in a series of steps. For example, formation of transient hydrogen-oxygen complexes for a short period of time, accompanied by partial transfer of hydrogen kinetic energy to oxygen eventually lead to capturing of hydrogen by oxygen and forming OH on the surface of GO.

Moreover, in regions of GO that the density of oxygen is high and they form cluster of oxygen, harmful chemical pathways by hydrogen can be dominant, in particular in high temperatures. We can single out a process in which a hydrogen molecule dissociated by two neighboring oxygen atoms and form a pair of hydroxyl groups close to each other.
Because this chemical reaction involves the energy transfer from hydrogen molecules to newly formed hydroxyl groups, the underlying C-C bonds undergo strong fluctuations that increases the chance of C-C bond breaking.
With increasing the density of hydrogen gas, oxygen localized on GO are nocked off with higher frequency allowing conversion of a pair of neighboring epoxys to carbonyl-holes and/or to hydroxyl-holes by collective C-C bond breaking.
This implies an optimal limit on density of hydrogen gas to achieve the highest quality for graphene samples.

\begin{figure}
\begin{center}
\includegraphics[width=0.5\linewidth]{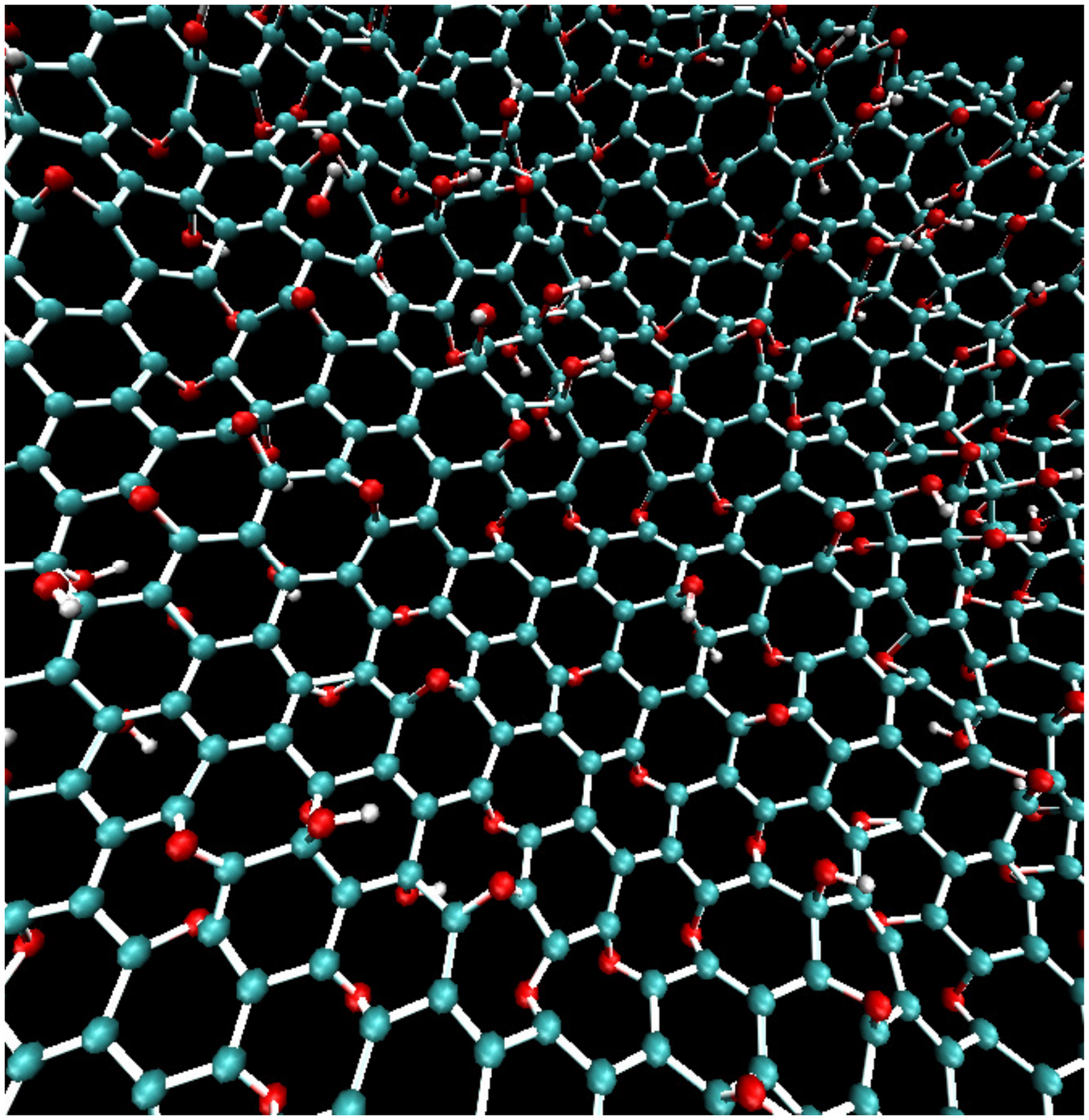}
\includegraphics[width=0.5\linewidth]{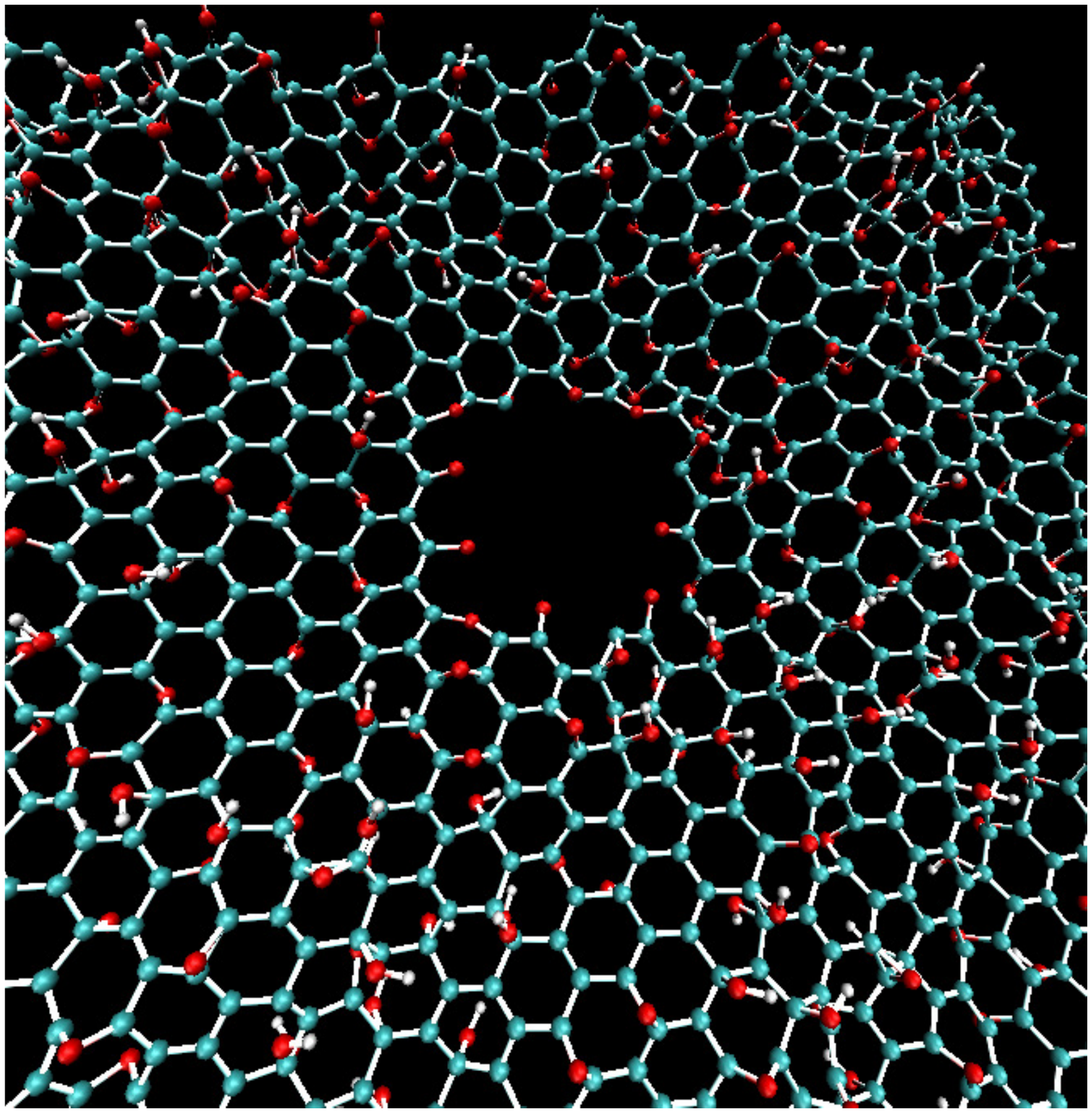}\\ \vspace{0.5cm}
\noindent
\caption{
Initial structure of relaxed GO (top) and GO with a large etch-hole (bottom) used to study for the modeling of thermal annealing in the hydrogen-rich environment. These structures were first relaxed to their equilibrium structure then used for MD calculation.
}
\label{fig3}
\end{center}\vspace{-0.5cm}
\end{figure}

We now turn to study the annealing process of GO that contains a large etch-hole. We further compare the annealing of GO with and without etch-hole.
Figs.~\ref{fig3} shows the post-optimized initial structure of a single sheet of GO and the one with a large etch-hole.
Unlike the previous calculation, the underlying structure of host-carbons is not a perfect periodic structure of graphene (because of structural optimization).
For a comparison, we build a similar structure for GO (without etch-hole) and perform similar post-optimization before starting thermal annealing. Here GO consists of $N_C=864$, $N_{\rm O}=255$, and $N_{\rm OH}=144$ corresponding to the number of carbon, oxygen (in form of mixture of carbonyl and epoxy groups) and hydroxyl-groups in GO, hence the oxygen percentage coverage is roughly 30\%.
For a GO with an etch-hole $N_C=836$. To passivate the sp$^2$ carbon-dangling bonds in the etch-hole, oxygen-atoms and hydroxyl groups are added hence the oxygen percentage coverage is slightly more than 30\%.
Moreover, the average number of carbon-oxygen bonds per oxygen $\langle N_{\rm C-O}\rangle\approx 1.6$ and $\langle N_{\rm C-O}\rangle\approx 1.5$ for GO and GO with the etch-hole. This is because in the etch-hole the oxygen-carbon form double-bonds.
The computational box is filled up to 2000 hydrogen-atoms. Its planar and perpendicular dimensions are $(4.3\times 5)\times 2$ nm$^3$ with $n_{\rm H}\approx 0.48 \times 10^{23} cm^{-3}$ denoting the maximum density of hydrogen gas considered in our simulation.

\begin{figure}
\begin{center}
\includegraphics[width=0.9\linewidth]{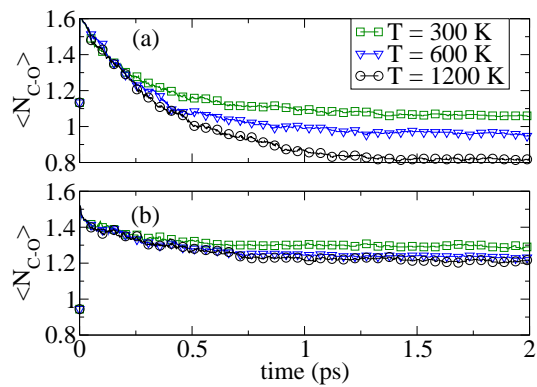}\\ \vspace{0.5cm}
\noindent
\caption{
The time evolution of average carbon-oxygon bonds as a function of temperature and initial hydrogen density of $n_{\rm H}\approx 0.24 \times 10^{23} {\rm cm}^{-3}$ for (a) GO and (b) GO with a single etch-hole.
}
\label{fig4}
\end{center}\vspace{-0.5cm}
\end{figure}

The time evolution of the oxygen concentration in GO as a function of temperature, measured by the average C-O bonds, is shown in Fig.~\ref{fig4}. As expected, the number of reduced oxygen increases with temperature.
Comparing the time evolution of oxygen removed from GO and GO with a large etch-hole shown in Fig.~\ref{fig4} (a) and (b), reveals that the initial oxygen abstraction close to large etch-holes take place with slower rate and without significant dependence on temperature, most likely because oxygen near etch-hole are more strongly bounded to GO than oxygen in the bulk~\cite{Gao2010:JPCC}.
It is interesting to note that the dynamical behavior of the etch-hole interacting with hydrogen-gas is different in high temperatures, e.g., close to $T=1500$ K the nucleation and growth of damaged C-C bonds in form of microscopic-cracks propagating toward the bulk is emerged. MD trajectories obtained by simulations of the etch-holes in vacuum and in the presence of hydrogen gas at $T=1500$ K are presented in on-line supplementary materials. To investigate the stability of etch-hole in lower temperatures, a series of simulation in hydrogen gas at $T=1200$ K with concentration up to
$n_{\rm H}\approx 0.8 \times 10^{23} cm^{-3}$ (up to two times larger than the corresponding hydrogen gas concentration used at $T=1500$ K) have been performed and no major damage in C-C bonds (as found at $T=1500$ K) observed.


\begin{figure}
\begin{center}
\includegraphics[width=0.9\linewidth]{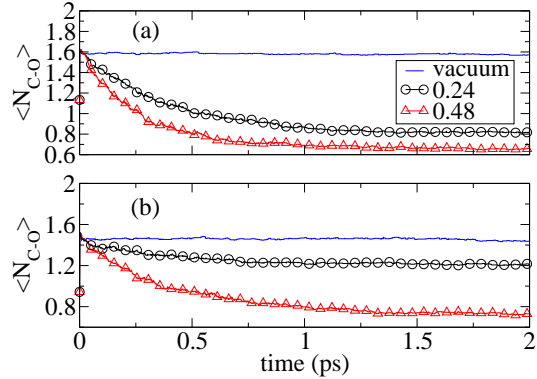}\\ \vspace{0.5cm}
\noindent
\caption{
The annealing at $T=1200$K as a function of density of $H_2$ molecules shown in the figure in unit of $10^{23}{\rm cm}^{-3}$ for (a) GO and (b) GO with a single etch-hole.
}
\label{fig5}
\end{center}\vspace{-0.5cm}
\end{figure}

\begin{figure}
\begin{center}
\includegraphics[width=0.9\linewidth]{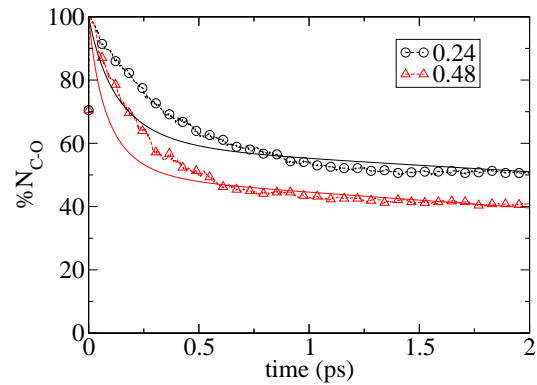}\\ \vspace{0.5cm}
\noindent
\caption{
The percentage oxygen remaining on GO after annealing at $T=1200$K as a function of density of $H_2$ molecules shown in the figure in unit of $10^{23}{\rm cm}^{-3}$ for a GO with no etch-hole. The bold-lines are the result of fitting of MD (the dashed-lines with open circles and triangles) to the mass-action equations~(\ref{eq2})-(\ref{eq3}) with $\alpha=0$, $k_1=0.15$ and $k_2=0.1$ ps$^{-1}$.
}
\label{fig9}
\end{center}\vspace{-0.5cm}
\end{figure}

The time-evolution of H-O bonds is governed by non-linear reaction-diffusion process. In a chemical kinetics model, mass-action equations describe the interaction of hydrogen molecules with GO to form complex, ${\rm C}^*$. This step is followed by a second-step of formation of a reaction product, P, and the release of new form of GO that contains less oxygen functional group
\begin{equation}
{\rm [GO] + [H] \rightarrow [C^*] \rightarrow [GO'] + [P]},
\label{eq1}
\end{equation}
with the reaction rates $k_1$ and $k_2$ associated with the first and second steps of chemical reactions. In this model we ignore the reverse reaction of ${\rm C^* \rightarrow GO + H_2}$, assuming $k_{-1} << k_2$. Also we assume that the kinetic model is reaction dominant such that the diffusion of hydrogen molecules occurs much faster than chemical reactions, hence the time-evolution of reaction-diffusion process is dominated by reaction rates.
Denoting $n_{\rm GO}$ and $n_{\rm GO'}$ as the initial and final concentration of oxygen in GO, and $n_{\rm C^*}$ and $n_{\rm P}$ as the concentration of all complexes and products,
the mass-action equations corresponding to the reaction of equation~\ref{eq1} are given by
\begin{eqnarray}
&&\frac{dn_{\rm GO}}{dt} = \alpha \frac{dn_{\rm GO}}{dt} - k_1 n_{\rm H}n_{\rm GO}, \nonumber \\ &&
\frac{dn_{\rm C^*}}{dt} = k_1 n_{\rm H}n_{\rm GO} - k_2 n_{\rm C^*},
\label{eq2}
\end{eqnarray}
subjected to the mass conservation $n_{\rm H} + n_{\rm C^*} = n_0$ where $n_0$ is the concentration of hydrogen molecules at $t=0$. Here $\alpha$ is the oxygenation rate of graphene that lead formation of epoxy and hydroxyl functional groups on GO.
To calculate solutions of Eqs.~\ref{eq1}-\ref{eq2} numerically we introduce scaled variables $g(t)=n_{\rm GO}(t)/n_0$, $c(t)=n_{\rm C^*}(t)/n_0$ and $r_1=k_1 n_0$. Using these new variables, Eqs.~\ref{eq2} transform to
\begin{eqnarray}
&&\frac{dg}{dt} =  - r_1 [1-c(t)]g(t), \nonumber \\ &&
\frac{dc}{dt} = r_1 [1-c(t)]g(t) - k_2 c(t).
\label{eq3}
\end{eqnarray}
The numerical solutions of Eqs.~\ref{eq1}-\ref{eq3} provide a fitting model to our MD simulation shown in Figs.~\ref{fig4}-\ref{fig5}.
Fig.~\ref{fig9} shows the percentage oxygen remaining on GO and the fitting to the mass-action equations with $k_1=0.15$ and $k_2=0.1$ ps$^{-1}$ after annealing ($\alpha=0$) at $T=1200$K for GO (with no hole).


\section{conclusion}
Ab-initio CPMD and ReaxFF molecular dynamic calculations performed in this work reveal that annealing of graphene-oxide in low-temperatures allows prevention of the structural damages due to relaxation of the lattice structure assisted by the oxygen surface diffusion. Our analysis indicates that the migration of epoxy functional groups provide a protection mechanism for C-sp$^2$ bonds hence preserving the integrity of the graphene honeycomb structure.
We further demonstrate that a chemically active environment, e.g. enriched with hydrogen gas, can assist removal of oxygen and blocks formation of carbonyl pairs, hence hindering the structural damage of GO either because of physical pressure exerted by the gas or imposed by chemical reactions.
Our analysis shows that hydrogen in environment do not suppress C-C bond breaking, but that H facilitate continuous removal of oxygen leading to reduced C-O bonding in GO, consistent with the expectation on the role of hydrogen in GO reduction.
Finally, the results obtained in our studies can be used as a guideline for identifying the experimental conditions that allows minimization of the mechanical and chemical damages to graphene and optimization of their performance.

\section{acknowledgement}
Authors would like to thank Dr. Yves Chabal, Dr. Alexandre Fonseca and Dr. Muge Acik for useful comments and discussion.
We thank the support from Texas Advanced Computing Center (TACC) for computer resources.




\end{document}